\documentclass[aps,pra,preprint,showpacs,groupedaddress]{revtex4}
\usepackage[dvips]{graphicx}
\usepackage{epsfig}
\usepackage{graphicx}
\usepackage{subfigure}
\def\be{\begin{equation}}
\def\ee{\end{equation}}
\def\ba{\begin{eqnarray}}
\def\ea{\end{eqnarray}}

\begin{document}
\title{Entanglement dynamics of one-dimensional driven spin systems in time-varying magnetic fields}
\author{Bedoor Alkurtass$^{1}$, Gehad Sadiek$^{1,2,}$\footnote{Corresponding author: gehad@ksu.edu.sa}, Sabre Kais$^{3}$}
\affiliation{
${}^1$ Department of Physics, King Saud University, Riyadh 11451, Saudi Arabia\\
${}^2$ Department of Physics, Ain Shams University, Cairo 11566, Egypt\\
${}^3$ Department of Chemistry and Birck Nanotechnology center, Purdue University, West Lafayette, Indiana 47907, USA\\}
\begin{abstract}
We study the dynamics of entanglement for a one-dimensional spin chain with a nearest neighbor time-dependent Heisenberg coupling $J(t)$ between the spins in presence of a time-dependent external magnetic field $h(t)$ at zero and finite temperatures. We consider different forms of time dependence for the coupling and magnetic field; exponential, hyperbolic and periodic. We examined the system size effect on the entanglement asymptotic value. It was found that for a small system size the entanglement starts to fluctuate within a short period of time after applying the time dependent coupling. The period of time increases as the system size increases and disappears completely as the size goes to infinity. We also found that when $J(t)$ is periodic the entanglement shows a periodic behavior with the same period, which disappears upon applying periodic magnetic field with the same frequency. Solving the particular case where J(t) and h(t) are proportional exactly, we showed that the asymptotic value of entanglement depends only on the initial conditions regardless of the form of $J(t)$ and $h(t)$ applied at $t>0$.
\end{abstract}
\maketitle
\section{Introduction}
Quantum entanglement represents one of the corner stones of the quantum mechanics theory with no classical analog \cite{Peres}. Quantum entanglement is a nonlocal correlation between two (or more) quantum systems such that the description of their states has to be done with reference to each other even if they are spatially well separated. Understanding and quantifying entanglement may provide an answer for many questions regarding the behavior of the many body quantum systems. Particularly, entanglement is considered as the physical property responsible for the long-range quantum correlations accompanying a quantum phase transition in many-body systems at zero temperature \cite{RMP,Osborne-QPT,Zhang-criticality}. 
Entanglement plays a crucial role in many fields of modern physics, particularly, quantum teleportation, quantum cryptography and quantum computing \cite{Nielsen, Boumeester}. It is considered as the physical basis for manipulating linear superpositions of quantum states to implement the different proposed quantum computing algorithms. Different physical systems have been proposed as promising candidates for the future quantum computing technology \cite{Barenco,Vandersypen,Chuang,Jones,Cirac,Monroe,Turchette,Averin,Shnirman}. It is a major task in each one of these considered systems 
to find a controllable mechanism to form and coherently manipulate the entanglement between a two-qubit system, creating an efficient quantum computing gate. The coherent manipulation of entangled states has been observed in different systems such as isolated trapped ions \cite{Chiaverini}, superconducting junctions \cite{Vion} and coupled quantum dots where the coupling mechanism in the latter system is the Heisenberg exchange interaction between electron spins \cite{Johnson,Koppens,Petta}. One of the most interesting proposals for creating a controllable mechanisms in coupled quantum dot systems was introduced by D. Loss et al. \cite{Spin_QD1,Spin_QD2}. The coupling mechanism is a time-dependent exchange interaction between the two valence spins on a doubled quantum dot system, which can be pulsed over definite intervals resulting a swap gate. This control can be achieved by raising and lowering the potential barrier between the two dots through controllable gate voltage. In a previous work, a two-atom system with time dependent coupling was studied and the critical dependence of the entanglement and variance squeezing on the strength and frequency of the coupling was demonstrated \cite{Abdalla}.

Quantifying entanglement in the quantum states of multiparticle systems is in the focus of interest in the field of quantum information.
However, quantum entanglement is very fragile due to the induced decoherence caused by the inevitable coupling to the environment. Decoherence is considered as one of the main obstacles toward realizing an effective quantum computing system \cite{Zurek1991}. The main effect of decoherence is to randomize the relative phases of the possible states of the considered system. Quantum error correction \cite{Shor1995} and decoherence free subspace \cite{Bacon2000,Divincenzo2000} have been proposed to protect the quantum property during the computation process. Nevertheless, offering a potentially ideal protection against environmentally induced decoherence is a difficult task. Moreover, a spin-pair entanglement is a reasonable measure for decoherence between the considered two-spin system and its environment constituted by the rest of spins on the chain. The coupling between the system and its environment leads to decoherence in the system and sweeping out entanglement between the two spins. Therefore, monitoring the entanglement dynamics in the considered system helps us to understand the behavior of the decoherence between the considered two spins and their environment. Particularly, the effect of the environment size on the coherence of quantum states of the system can be considered by watching the spin pair entanglement evolution versus the the number of sites $N$ in the chain.

Developing new experimental techniques enabled the generation and control of multiparticle entanglement \cite{Eibl_03,Wieczorek_08,R�dmark_09,Prevedel_09,Barz_10,Krischek_10} as well as the fabrication of one dimensional spin chains \cite{ZMWang_04,Tenga_07,Ugur_09}. This progress in the experimental arena sparked an intensive theoretical research over the multiparticle systems and particularly the one dimensional spin chains \cite{Rossignoli_05,Huang_06,Rossini_07,Furman_08,Skrovseth_09,Physica_B_404, Sadiek_Nuovo_2010,Burrell_09,Ren_10, Niederberger_10}. The dynamics of entanglement in an $XY$ and Ising spin chains has been studied considering a constant nearest neighbor exchange interaction, in presence of a time varying magnetic field represented by a step, exponential and sinusoidal functions of time \cite{HuangQInfo,HuangPhysRev}. Furthermore, the dynamics of entanglement in a one dimensional Ising spin chain at zero temperature was investigated numerically where the number of spins was seven at most \cite{Dyn_Ising}. The generation and transportation of the entanglement through the chain, which irradiated by a weak resonant field under the effect of an external magnetic field were investigated. Recently, the entanglement in anisotropic $XY$ model with a small number of spins, with a time dependent nearest neighbor coupling at zero temperature was studied too \cite{Driven_xy_model}. The time-dependent spin-spin coupling was represented by a dc part and a sinusoidal ac part. It was observed that there is an entanglement resonance through the chain whenever the ac coupling frequency is matching the Zeeman splitting. Very recently, we have studied the time evolution of entanglement in a one dimensional spin chain in presence of a time dependent magnetic field $h(t)$ considering a time dependent coupling parameter $J(t)$ where both $h(t)$ and $J(t)$ were assumed to be of a step function form \cite{Sadiek_2010}. Solving the problem exactly, we found that the system undergoes a nonergodic behavior. At zero temperature we found that the asymptotic value of the entanglement depends only on the ratio $\lambda= J/h$. However, at nonzero temperatures it depends on the individual values of $h$ and $J$. Also we have demonstrated that the quantum effects dominate within certain regions of the temperature-$\lambda$ space that vary significantly depending on the degree of the anisotropy of the system.

In this work, we investigate the time evolution of quantum entanglement in a one dimensional $XY$ spin chain system coupled through nearest neighbor interaction under the effect of an external magnetic field at zero and finite temperature. We consider both time-dependent nearest neighbor Heisenberg coupling $J(t)$ between the spins on the chain and magnetic field $h(t)$, where the function forms are exponential, periodic and hyperbolic in time.

This paper is organized as follows. In Sec. II, we present our model and discuss the numerical solution for the the $XY$ spin chain for a general form of the coupling and magnetic field. Then, we present an exact solution for the system for the special case $J(t)= \lambda h(t)$, where $\lambda$ is a constant. In Sec. III, we evaluate the entanglement using the magnetization and the spin-spin correlation functions of the system. We present our results and discuss them in sec. IV. Finally, in Sec. V we conclude and discuss future directions.
\section{THE TIME DEPENDENT XY MODEL}
\subsection{A Numerical Solution}
In this section, we present a numerical solution for the $XY$ model of a spin chain with $N$ sites in the presence of a time-dependent external magnetic field $h(t)$. We consider a time-dependent coupling $J(t)$ between the nearest neighbor spins on the chain. The Hamiltonian for such a system is given by

\begin{equation}
H=-\frac{J(t)}{2} (1+\gamma) \sum_{i=1}^{N} \sigma_{i}^{x} \sigma_{i+1}^{x}-\frac{J(t)}{2}(1-\gamma)\sum_{i=1}^{N} \sigma_{i}^{y} \sigma_{i+1}^{y}- \sum_{i=1}^{N} h(t) \sigma_{i}^{z}\, ,
\label{eq:H}
\end{equation}
where $\sigma_{i}$'s are the Pauli matrices and $\gamma$ is the anisotropy parameter. For simplicity, we'll consider $\hbar=1$ throughout this paper.
Defining the raising and lowering operators $a^{\dagger}_{i}$, $a_{i}$
\begin{equation}
a^{\dagger}_{i} = \frac{1}{2} (\sigma_{i}^{x}+ i \sigma_{i}^{y}), \;\;\; a_{i} = \frac{1}{2} (\sigma_{i}^{x}- i \sigma_{i}^{y})\, .
\label{eq:raisinglowering}\end{equation}
Following the standard procedure to treat the Hamiltonian (\ref{eq:H}), we introduce Fermi operators $b^{\dagger}_{i}$, $b_{i}$ \cite{LSM}
\begin{equation}
a_{i}^{\dagger}=b_{i}^{\dagger} \exp(i \pi \sum_{j=1}^{i-1}b_{j}^{\dagger}b_{j}), \;\;\; a_{i}= \exp(-i \pi \sum_{j=1}^{i-1}b_{j}^{\dagger}b_{j})b_{i}\, ,
\label{eq:fraisinglowering}\end{equation}
then applying Fourier transformation we obtain
\begin{equation}
b^{\dagger}_{i} = \frac{1}{\sqrt{N}} \sum_{p=-N/2}^{N/2} e^{i j \phi_{p}} c^{\dagger}_{p}, \;\;\; b_{i} = \frac{1}{\sqrt{N}} \sum_{p=-N/2}^{N/2} e^{-i j \phi_{p}} c_{p}\, .
\label{eq:fourierraisinglowering}\end{equation}
where $\phi_{p}=\frac{2 \pi p}{N}$. Therefore, the Hamiltonian can be written as
\begin{equation}
H=\sum_{p=1}^{N/2} \tilde{H}_{p}\, ,
\label{eq:Hsum}\end{equation}
with $\tilde{H}_{p}$ given by
\begin{equation}
\tilde{H}_{p}=\alpha_{p}(t) [c_{p}^{\dagger} c_{p}+c_{-p}^{\dagger} c_{-p}]+i J(t) \delta_{p} [c_{p}^{\dagger} c_{-p}^{\dagger}+c_{p} c_{-p}]+2 h(t)\, ,
\label{eq:Hp}\end{equation}
where $\alpha_{p}(t)=-2 J(t) \cos \phi_{p} - 2 h(t)$ and  $\delta_{p}=2 \gamma \sin \phi_{p}$.

As $[\tilde{H}_{l},\tilde{H}_{m}]=0$ for $l,m=0,1,2,\dots,N/2$, the Hamiltonian in the $2^N$-dimensional Hilbert space can be decomposed into $N/2$ non-commuting sub-Hamiltonians, each in a 4-dimensional independent subspace.
Using the basis $\{ \left|0\right\rangle, c_{p}^{\dagger}c_{-p}^{\dagger}\left|0\right\rangle, c_{p}^{\dagger}\left|0\right\rangle, c_{-p}^{\dagger}\left|0\right\rangle \}$ we obtain the matrix representation of $\tilde{H}_{p}$
\begin{equation}
\tilde{H}_{p}=\left(\begin{array} {cccc}
2 h(t) & -i J(t)\delta_{p} & 0 & 0\\
i J(t) \delta_{p} & -4 J(t)\cos \phi_{p}-2 h(t) & 0 & 0\\
0 & 0 & -2 J(t)\cos \phi_{p} & 0\\
0 & 0 & 0 & -2 J(t)\cos \phi_{p}\\
\end{array}\right)\, .
\label{eq:Hmatrix}\end{equation}

Initially the system is assumed to be in a thermal equilibrium state and therefore its initial density matrix is given by
\begin{equation}
\rho_{p}(0)=e^{-\beta \tilde{H}_{p}(0)}\, ,
\label{eq:rho0}\end{equation}
where $\beta=1/k T$, $k$ is Boltzmann constant and $T$ is the temperature.

Since the Hamiltonian is decomposable we can find the density matrix at any time $t$, $\rho_{p}(t)$, for the $p$th subspace by solving Liouville equation given by
\begin{equation}
i \dot{\rho}_{p}(t)=[\tilde{H}_p(t),\rho_{p}(t)] \, ,
\label{eq:Liouville}
\end{equation}
which gives
\begin{equation}
\rho_{p}(t)=U_{p}(t) \rho_{p}(0) U_{p}^{\dagger}(t) \, .
\label{eq:UrhoU}
\end{equation}
where $U_{p}(t)$ is time evolution matrix which can be obtained by solving the equation
\begin{equation}
i \: \dot{U}_{p}(t)=U_{p}(t) \tilde{H}_{p}(t) \, .
\label{eq:Udot}
\end{equation}

To study the effect of a time-varying coupling parameter $J(t)$ we consider the following forms
\begin{eqnarray}
J_{exp}(t)&=&J_{1}+\left(J_{0}-J_{1}\right) e^{-K t} \, ,\\
J_{cos}(t)&=&J_{0}-J_{0} \cos\left(K t\right)\, ,\quad\quad\\
J_{sin}(t)&=&J_{0}-J_{0} \sin\left(K t\right)\, ,\quad\quad\\
J_{tanh}(t)&=&J_{0}+\frac{J_1-J_0}{2} \left[\tanh\left(K (t-\frac{5}{2})\right)+1\right]\,.
\end{eqnarray}

Note that Eq.~(\ref{eq:Udot}) gives two systems of coupled differential equations with variable coefficients. Such systems can only be solved numerically which we adopt in this paper.
\subsection{An Exact Solution for Proportional $J$ and $h$}

In this section we present an exact solution of the system using a general time-dependent coupling $J(t)$ and a magnetic field with the following form:
\be
J(t)=\lambda \: h(t)
\label{eq:proportional}
\ee
where $\lambda$ is a constant. Using Eqs.~(\ref{eq:Hmatrix}), (\ref{eq:Udot}) and (\ref{eq:proportional}) we obtain
\be
i \left(\begin{array}{cc}
\dot{u}_{11} & \dot{u}_{12}\\
\dot{u}_{21} & \dot{u}_{22}\\
\end{array}\right)=	\left(\begin{array}{cc}
u_{11} & u_{12}\\
u_{21} & u_{22}\\
\end{array}\right) \left(\begin{array}{cc}
\frac{2}{\lambda} & -i \delta_p\\
i \delta_p & -4 \cos \phi_p - \frac{2}{\lambda}\\
\end{array}\right) J(t)\, ,
\label{eq:U2dot}
\ee
and
\be
i \: \dot{u}_{33}=-2 \: \cos \phi_p \: J(t) \: u_{33}\, , \, \, \, \, \, u_{44}=u_{33}\, .
\ee

Equation (\ref{eq:U2dot}) can be rewritten as 
\be
i \: \dot{u}_j = J(t) \: H' u_j \, .
\label{DE}
\ee
for $j=1,2$, where 
\be
H'= \left(\begin{array}{cc}
\frac{2}{\lambda} & i \delta_p\\
-i \delta_p & 4\cos\phi_p-\frac{2}{\lambda}\\
\end{array}\right)\! , \: \: \: u_j=\left(\begin{array}{c}
u_{j1}\\
u_{j2}\\
\end{array}\right)\, .
\ee
Introducing a unitary rotation matrix
\be
S=\left(\begin{array}{cc}
\cos\theta & e^{i\phi} \sin\theta\\
-e^{-i\phi} \sin\theta & \cos\theta\\
\end{array}\right) \, .
\ee
Using $S$ to diagonalize $H'$ we obtain
\be
S H' S^{-1}=\left(\begin{array}{cc}
\lambda_1 & 0\\
0 & \lambda_2\\
\end{array}\right)\, .
\ee
Where the angles $\phi$ and $\theta$ were found to be
\be
\phi=(n+1) \pi, \: \:\:\: \tan{2\theta} = \frac{\delta_p}{2\cos\phi_p+\frac{2}{\lambda}}\: ,
\ee
where $n=0,\pm1,\pm2,\ldots$, therefore
\be
\sin{2\theta} = \frac{\delta_p}{\sqrt{\delta_p^2+(2\cos\phi_p+\frac{2}{\lambda})}}, \, \:\:\: \cos{2\theta} = \frac{2\cos\phi_p+\frac{2}{\lambda}}{\sqrt{\delta_p^2+(2\cos\phi_p+\frac{2}{\lambda})}}\,.
\ee
Finding $\lambda_1$ and $\lambda_2$ we get
\be
\lambda_1=\sqrt{\delta_p^2+(2\cos\phi_p+\frac{2}{\lambda})}-2\cos\phi_p, \, \lambda_2=-\sqrt{\delta_p^2+(2\cos\phi_p+\frac{2}{\lambda})}-2\cos\phi_p\, .
\ee
Now we define $v_j=S u_j$ and substitute in eq.~(\ref{DE}) we get
\be
i \: \dot{v_j}=\left(S H' S^{-1} + i \dot{S} S^{-1}\right) v_j \, .
\ee
Hence
\be
i \: \dot{v_j}=\left(\begin{array}{cc}
\lambda_1 & 0\\
0 & \lambda_2\\
\end{array}\right) v_j\,.
\ee
Solving this equation we obtain
\be
v_1=\left(\begin{array}{cc}
\cos \theta \: e^{-i \lambda_1\int^t_0{J(t') dt'}}\\
i\sin \theta \: e^{-i \lambda_2\int^t_0{J(t') dt'}}\\
\end{array}\right) ,\,v_2=\left(\begin{array}{cc}
i\sin \theta \: e^{-i \lambda_1\int^t_0{J(t') dt'}}\\
\cos \theta \: e^{-i \lambda_2\int^t_0{J(t') dt'}}\\
\end{array}\right) \,.
\ee
Finally $u$ is given by
\be
u_{11}=\cos^2 \theta e^{-i \lambda_1\int^t_0{J(t') dt'}} + \sin^2 \theta e^{-i \lambda_2\int^t_0{J(t') dt'}}\, ,
\ee
\be
u_{12}=-i\sin \theta\cos \theta \left\{e^{-i \lambda_1\int^t_0{J(t') dt'}} - e^{-i \lambda_2\int^t_0{J(t') dt'}}\right\}\, ,
\ee
\be
u_{21}=-u_{12}\, ,
\ee
\be
u_{22}=\sin^2 \theta e^{-i \lambda_1\int^t_0{J(t') dt'}} + \cos^2 \theta e^{-i \lambda_2\int^t_0{J(t') dt'}}\, ,
\ee
\be
u_{33}=u_{44}=e^{2i \cos\phi_p \int^t_0{J(t') dt'}} \, ,
\ee
where
\be
\sin\theta=\sqrt{\frac{\sqrt{\delta_p^2+(2\cos\phi_p+\frac{2}{\lambda})}-(2\cos\phi_p+\frac{2}{\lambda})}{2\sqrt{\delta_p^2+(2\cos\phi_p+\frac{2}{\lambda})}}}\, ,
\ee
\be
\cos\theta=\sqrt{\frac{\sqrt{\delta_p^2+(2\cos\phi_p+\frac{2}{\lambda})}+(2\cos\phi_p+\frac{2}{\lambda})}{2\sqrt{\delta_p^2+(2\cos\phi_p+\frac{2}{\lambda})}}}\, .
\ee
\section{Spin Correlation Functions and Entanglement Evaluation}
In this section we evaluate different magnetization and the spin-spin correlation functions of the $XY$ model, then we evaluate the entanglement in the system. The magnetization in the $z$-direction is defined as
\begin{equation}
M=\frac{1}{N}\sum_{j=1}^{N}(S_{j}^{z})=\frac{1}{N}\sum_{p=1}^{1/N}M_p \:,
\label{eq:Mdef}
\end{equation}
where $M_p=c_{p}^{\dagger} c_{p}+c_{-p}^{\dagger} c_{-p}-1$. In terms of the density matrix, it is given by
\begin{equation}
\left\langle M_{z}\right\rangle=\frac{Tr[M\rho(t)]}{Tr[\rho(t)]} = \frac{1}{N}\sum_{p=1}^{1/N}\frac{Tr[M_{p}\rho_{p}(t)]}{Tr[\rho_{p}(t)]} \,.
\label{eq:Mexp}
\end{equation}
The spin correlation functions are defined by
\begin{equation}
S^{x}_{l,m}=\left\langle S^{x}_{l} S^{x}_{m} \right\rangle, \;\;\;S^{y}_{l,m}=\left\langle S^{y}_{l} S^{y}_{m} \right\rangle, \;\;\; S^{z}_{l,m}=\left\langle S^{z}_{l} S^{z}_{m} \right\rangle\, ,
\label{eq:Sdef}\end{equation}
which can be written in terms of the fermionic operators as follows \cite{LSM}:

\begin{equation}
S_{l,m}^{x}=\frac{1}{4}\left\langle B_{l} A_{l+1} B_{l+1}\ldots A_{m-1} B_{m-1} A_{m}\right\rangle\, ,
\label{eq:Sxdef}\end{equation}

\begin{equation}
S_{l,m}^{y}=\frac{\left(-1\right)^{l-m}}{4}\left\langle A_{l} B_{l+1} A_{l+1}\ldots B_{m-1} A_{m-1} B_{m}\right\rangle\, ,
\label{eq:Sydef}\end{equation}

\begin{equation}
S_{l,m}^{z}=\frac{1}{4}\left\langle A_{l} B_{l} A_{m} B_{m}\right\rangle\, ,
\label{eq:Szdef}\end{equation}
where
\begin{equation}
A_{i}=b_{i}^{\dagger}+b_{i}, \;\;\;B_{i}=b_{i}^{\dagger}-b_{i}\, .
\label{eq:AandB}\end{equation}
Using Wick Theorem \cite{Wick}, the expressions (\ref{eq:Sxdef})-(\ref{eq:Szdef}) can be evaluated as pfaffians of the form
\begin{equation}
S_{l,m}^{x}=\frac{1}{4} pf \left(\begin{array}{cccccc}
0 &
 F_{l, l+1} & G_{l, l+1} & \cdots & G_{l, m-1} & F_{l, m}\\
 & 0 & P_{l+1, l+1} & \cdots & P_{l+1, m-1} & Q_{l+1, m}\\
 &  &  & \cdots & . & .\\
 &  &  & & P_{m-1, m-1} & Q_{m-1, m}\\
 &  &  & & 0 & F_{m-1, m}\\
 &  &  & &  & 0 \end{array}\right) \, ,
\label{eq:Sxpf}
\end{equation}
\begin{equation}
S_{l,m}^{y}=\frac{\left(-1\right)^{l-m}}{4} pf \left(\begin{array}{cccccc}
0 & P_{l, l+1} & Q_{l, l+1} & \cdots & Q_{l, m-1} & P_{l, m}\\
 & 0 & F_{l+1, l+1} & \cdots & F_{l+1, m-1} & G_{l+1, m}\\
 &  &  & \cdots & . & .\\
 &  &  & & F_{m-1, m-1} & G_{m-1, m}\\
 &  &  &  & 0 & P_{m-1, m}\\
 &  &  &  &  & 0 \end{array}\right) \, ,
\label{eq:Sybf}
\end{equation}
\begin{equation}
S_{l,m}^{z}=\frac{1}{4} pf \left(\begin{array}{cccc}
0 &  P_{l, l} &  Q_{l, m} &  P_{l, m}\\
 & 0 &  F_{l, m} &  G_{l, m}\\
 &  & 0 &  P_{m, m}\\
 &  &  & 0
\end{array}\right) \, ,
\label{eq:Szbf}\end{equation}
where
\begin{equation}
F_{l,m}=\left\langle B_{l} A_{m}\right\rangle, \;\;\; P_{l,m}=\left\langle A_{l} B_{m}\right\rangle,\;\;\; Q_{l,m}=\left\langle A_{l} A_{m}\right\rangle,\;\;\; G_{l,m}=\left\langle B_{l} B_{m}\right\rangle.
\end{equation}


To evaluate the entanglement between two quantum systems in the chain we use the concurrence which has been shown to be a measure of entanglement \cite{Wootters}. The concurrence $C(t)$ is defined as
\begin{equation}
C(\rho)=\max(0,\lambda_{a}-\lambda_{b}-\lambda_{c}-\lambda_{d})\, ,
\label{eq:C}
\end{equation}
where the $\lambda_{i}$'s are the positive square root of the eigenvalues, in a descending order, of the matrix $R$ defined by
\begin{equation}
R=\sqrt{\sqrt{\rho}\tilde{\rho}\sqrt{\rho}}\, ,
\label{eq:R}
\end{equation}
and $\tilde{\rho}$ is the spin-flipped density matrix given by
\begin{equation}
\tilde{\rho}=(\sigma_{y}\otimes\sigma_{y})\rho^{*}(\sigma_{y}\otimes\sigma_{y})\, .
\label{eq:tilderho}
\end{equation}
Knowing that $\rho$ is symmetrical and real due to the symmetries of the Hamiltonian and particularly the global phase flip symmetry, there will be only 6 non-zero distinguished matrix elements of $\rho$ which takes the form \cite{H_rho_symmetry}
\begin{equation}
\rho=\left(\begin{array}{cccc}
\rho_{1,1} & 0 & 0 & \rho_{1,4}\\
0 & \rho_{2,2} & \rho_{2,3} & 0\\
0 & \rho_{2,3} & \rho_{3,3} & 0\\
\rho_{1,4} & 0 & 0 & \rho_{4,4}\\
\end{array}\right)\, .
\label{eq:rhomatrix}
\end{equation}
Hence, the roots of the matrix $R$ come out to be $\lambda_{a}=\sqrt{\rho_{1,1}\rho_{4,4}}+\left|\rho_{1,4}\right|$, $\lambda_{b}=\sqrt{\rho_{2,2}\rho_{3,3}}+\left|\rho_{2,3}\right|$, $\lambda_{c}=\left|\sqrt{\rho_{1,1}\rho_{4,4}}-\left|\rho_{1,4}\right|\right|$ and $\lambda_{d}=\left|\sqrt{\rho_{2,2}\rho_{3,3}}-\left|\rho_{2,3}\right|\right|$.

To find the non-zero matrix elements of $\rho$, one can use the formula of the expectation value of an operator in terms of density matrix $\left\langle \hat{G} \right\rangle=Tr(\rho \: \hat{G})/ \: Tr(\rho)$ along with the magnetization eq.(\ref{eq:Mexp}) and the spin correlation functions eq.(\ref{eq:Sxdef})-(\ref{eq:Szdef}) which give
\begin{equation}\rho_{1,1}=\frac{1}{2} M^z_{l}+\frac{1}{2} M^z_{m}+S^z_{l,m}+\frac{1}{4}\, ,\label{eq:frho11}\end{equation}
\begin{equation}\rho_{2,2}=\frac{1}{2} M^z_{l}-\frac{1}{2} M^z_{m}-S^z_{l,m}+\frac{1}{4}\, ,\label{eq:frho22}\end{equation}
\begin{equation}\rho_{3,3}=\frac{1}{2} M^z_{m}-\frac{1}{2} M^z_{m}-S^z_{l,m}+\frac{1}{4}\, ,\label{eq:frho33}\end{equation}
\begin{equation}\rho_{4,4}=-\frac{1}{2} M^z_{l}-\frac{1}{2} M^z_{m}+S^z_{l,m}+\frac{1}{4}\, ,\label{eq:frho44}\end{equation}
\begin{equation}\rho_{2,3}=S^x_{l,m}+S^y_{l,m}\, ,\label{eq:frho23}\end{equation}
\begin{equation}\rho_{1,4}=S^x_{l,m}-S^y_{l,m}\, .\label{eq:frho14}\end{equation}
\section{Results and Discussion}

\subsection{Constant Magnetic Field}

\begin{figure}[htbp]
\begin{minipage}[c]{\textwidth}
 \centering 
   \subfigure[]{\label{fig:expK01}\includegraphics[width=7cm]{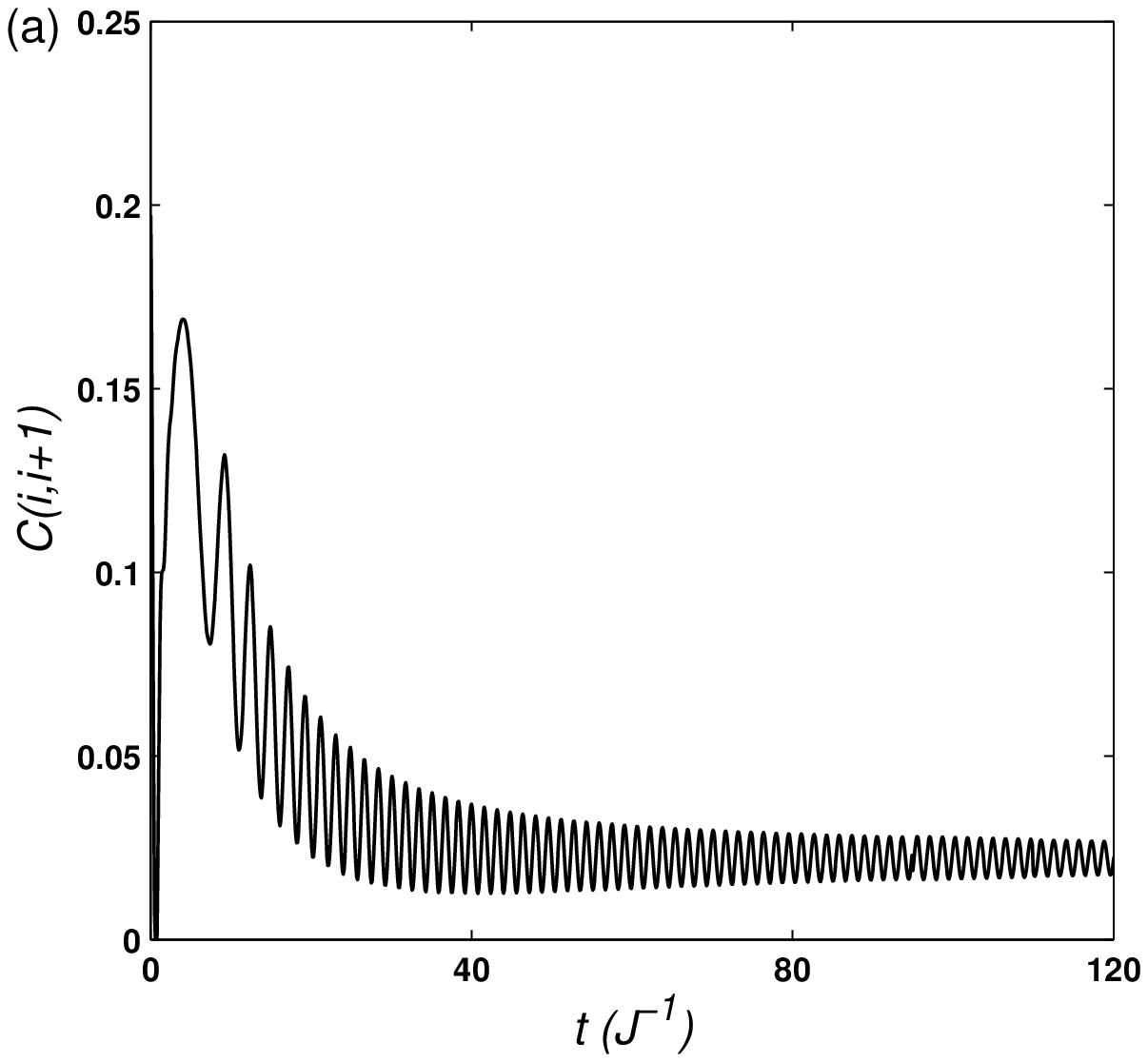}}\quad
   \subfigure[]{\label{fig:expK10}\includegraphics[width=7cm]{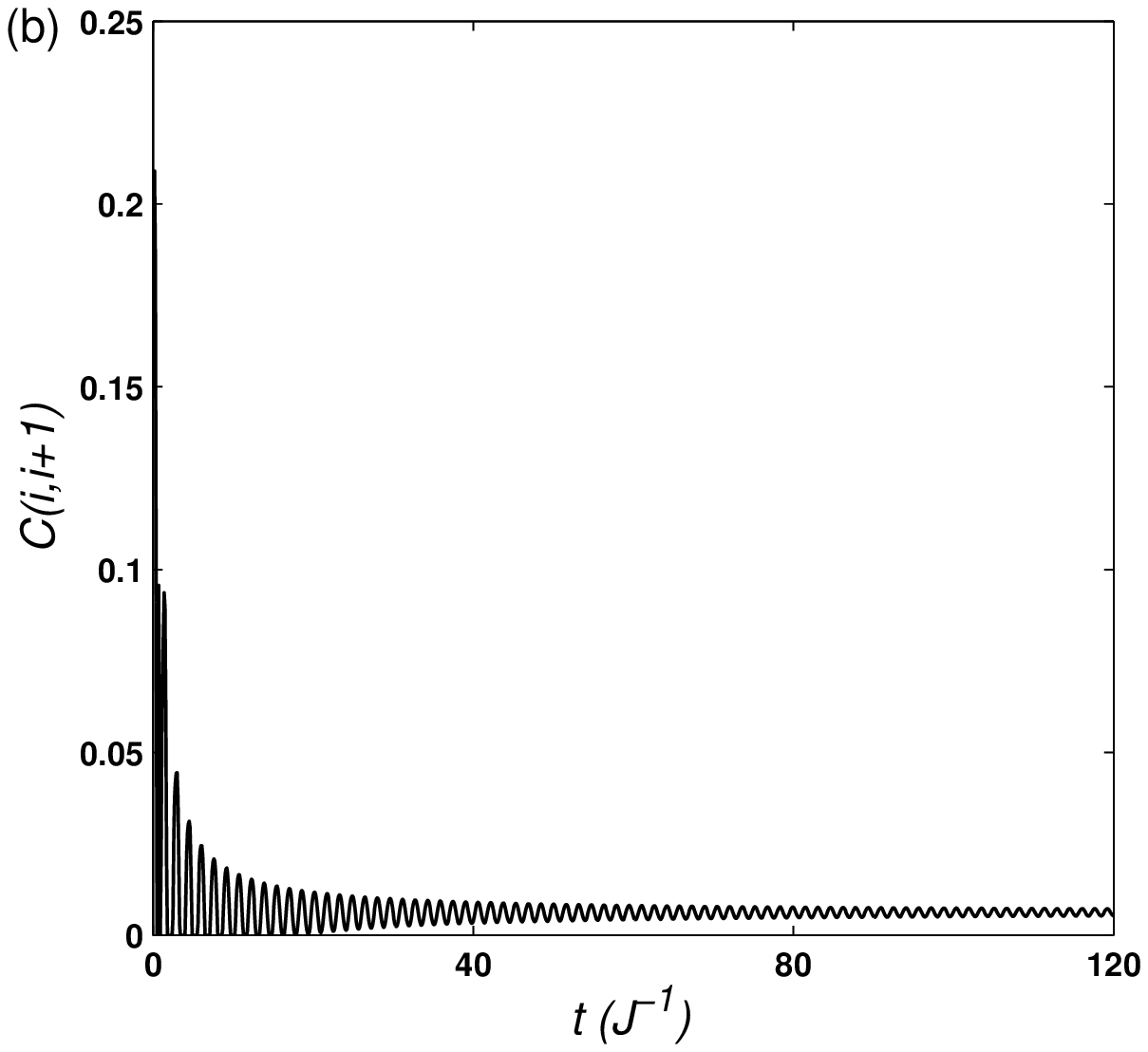}}\\
   \subfigure[]{\label{fig:tanhK01}\includegraphics[width=7cm]{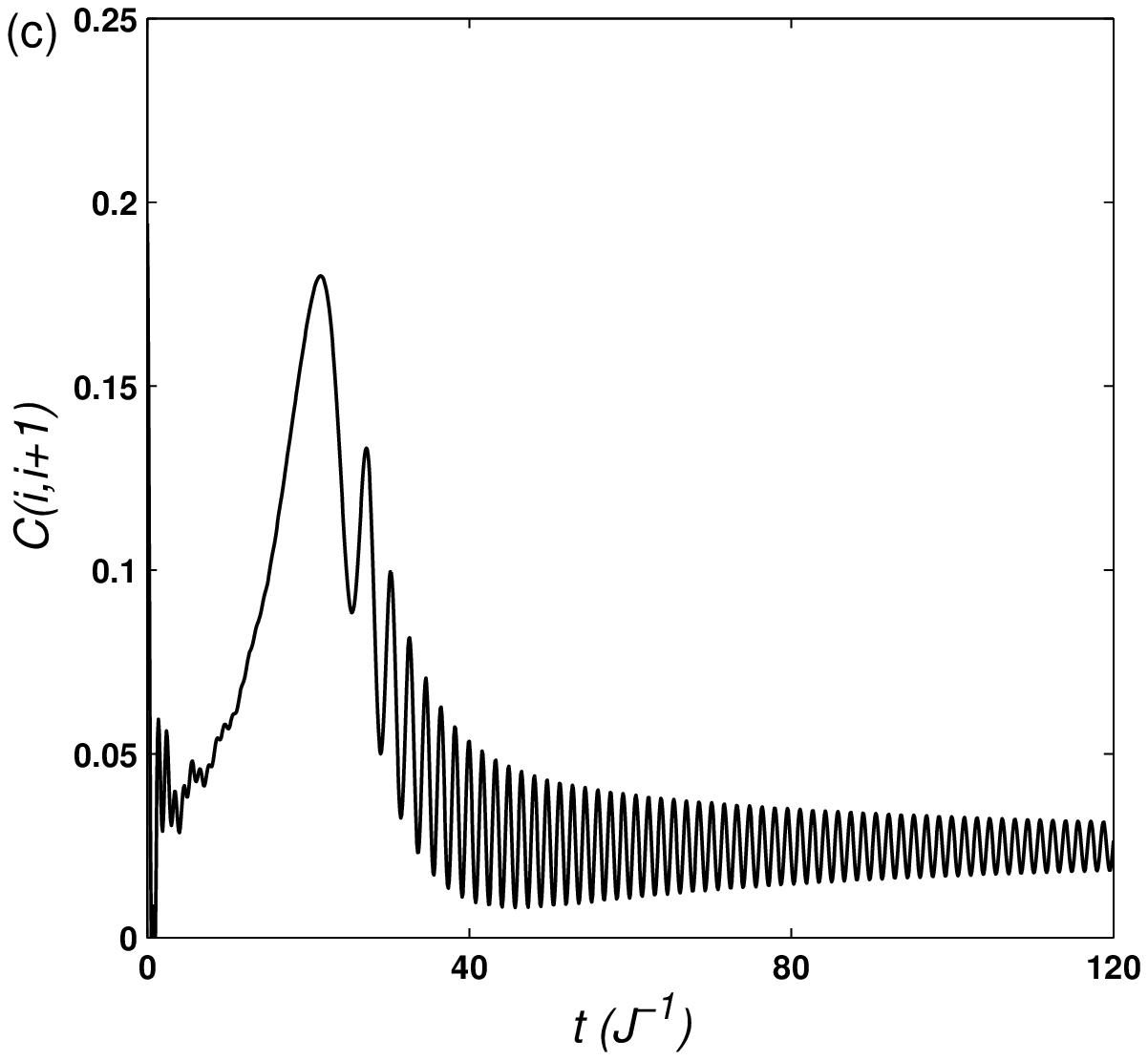}}\quad
   \subfigure[]{\label{fig:tanhK10}\includegraphics[width=7cm]{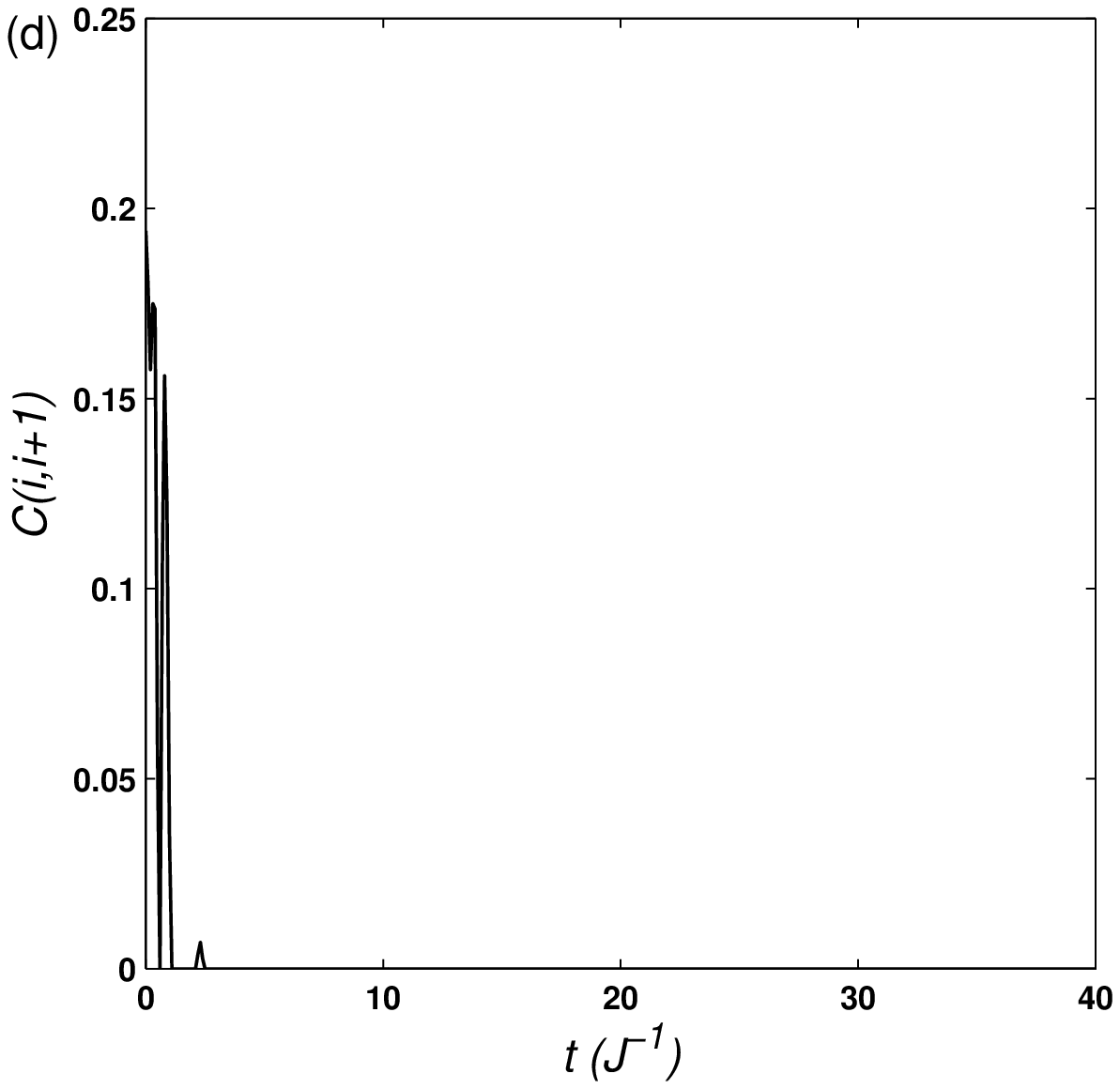}}\\
   \caption{{\protect\footnotesize $C(i,i+1)$ as a function of $t$ with $J_{0}=0.5, J_{1}=2, h=1, N=1000$ at $kT=0$ and (a) $J= J_{exp}, K=0.1$ (b) $J= J_{exp}, K=10$ ; (c) $J= J_{tanh}, K=0.1$ ; (d) $J= J_{tanh}, K=10$.}}
  \label{fig:Ks}
 \end{minipage}
\end{figure}

\begin{figure}[htbp]
 \centering 
	\includegraphics[width=14cm]{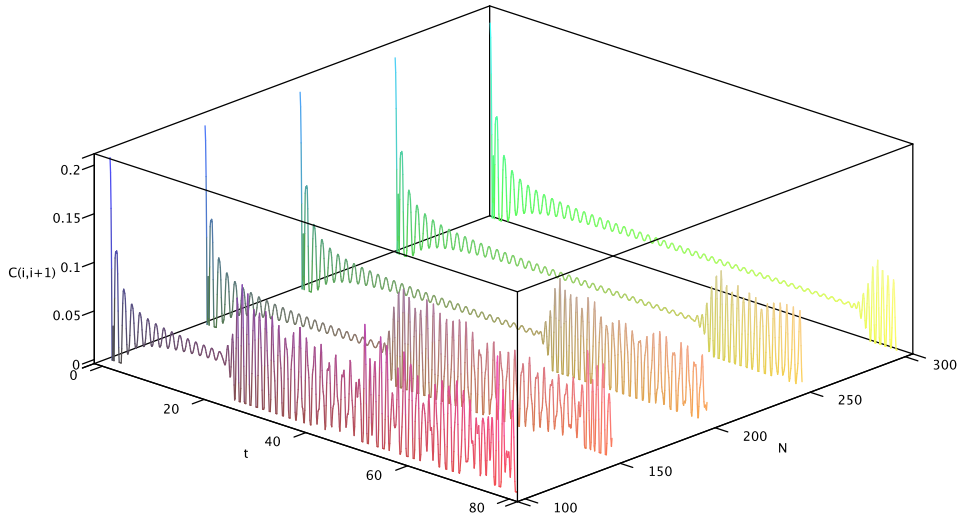}
   \caption{{\protect\footnotesize $C(i,i+1)$ as a function of $t$ (units of $J^{-1}$) with $J= J_{exp}, J_{0}=0.5, J_{1}=2, h=1, K=1000$ at $kT=0$ and $N$ varies from $100$ to $300$.}}
  \label{fig:Ns}
\end{figure}

\begin{figure}[tbph]
\begin{minipage}[c]{\textwidth}
 \centering 
   \subfigure[]{\label{fig:Jcosa}\includegraphics[width=6.5cm]{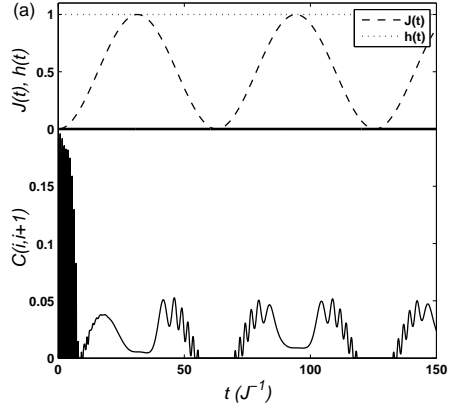}}\\
   \subfigure[]{\label{fig:Jcosb}\includegraphics[width=6.5cm]{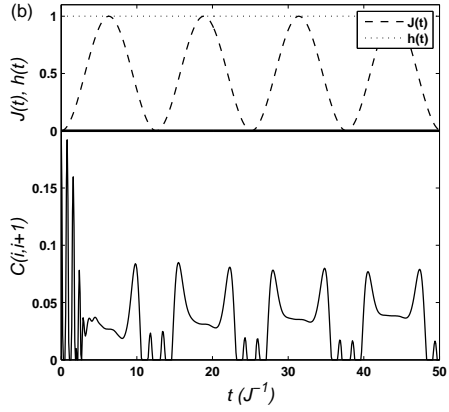}}\\
   \subfigure[]{\label{fig:Jcosc}\includegraphics[width=6.5cm]{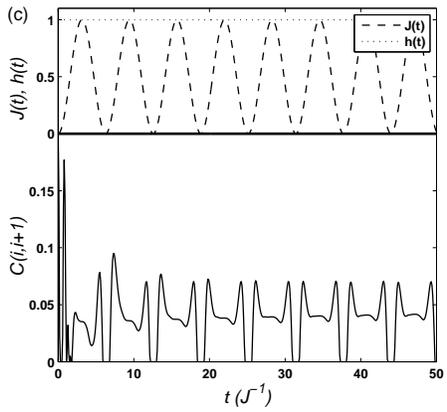}}
   \caption{{\protect\footnotesize Dynamics of nearest neighbor concurrence with $\gamma=1$ for $J_{cos}$ where $J_{0}=0.5, h=1$ at $kT=0$ and (a) $K=0.1$ ; (b) $K=0.5$ ; (c) $K=1$.}}
   \label{fig:Jcos}
   \end{minipage}
\end{figure}

\begin{figure}[tbph]
\begin{minipage}[c]{\textwidth}
 \centering 
   \subfigure[]{\label{fig:Jsina}\includegraphics[width=6.5cm]{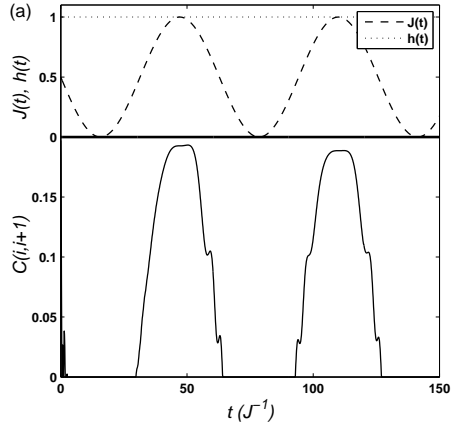}}\\
   \subfigure[]{\label{fig:Jsinb}\includegraphics[width=6.5cm]{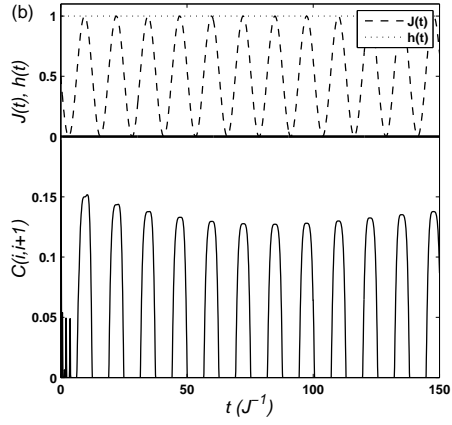}}\\
   \subfigure[]{\label{fig:Jsinc}\includegraphics[width=6.5cm]{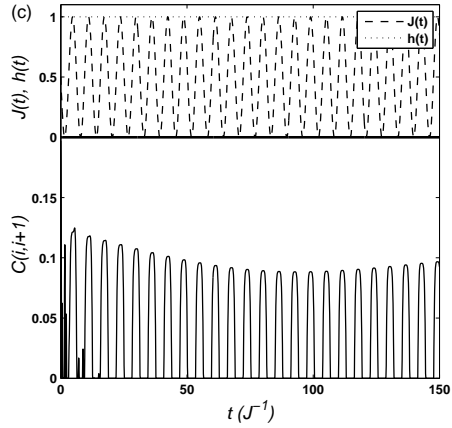}}
   \caption{{\protect\footnotesize Dynamics of nearest neighbor concurrence with $\gamma=1$ for $J_{sin}$ with $J_{0}=0.5, h=1$ at $kT=0$ and (a) $K=0.1$ ; (b) $K=0.5$ ; (c) $K=1$.}}
   \label{fig:Jsin}
   \end{minipage}
\end{figure}

We start with studying the dynamics of the nearest neighbor concurrence $C(i,i+1)$ for the completely anisotropic system, $\gamma=1$, when the coupling parameter is $J_{exp}$ as well as $J_{tanh}$ and the magnetic field is a constant using the numerical solution. In Figure~\ref{fig:Ks} we study the dynamics of the concurrence with the parameters $J_{0}=0.5, J_{1}=2, h=1$ and different values of the transition constant $K=0.1$ and 10. We note that the asymptotic value of the concurrence depends on $K$ in addition to the coupling parameter and magnetic field. The larger the transition constant is, the lower is the asymptotic value of the entanglement and the more rapid decay is. This result demonstrates the non-ergodic behavior of the system, where the asymptotic value of the entanglement is different from the one obtained under constant coupling $J_{1}$.
In Fig.~\ref{fig:Ns} we study the effect of the system size $N$ on the dynamics of the concurrence. We select the parameters $J_{0}=0.5, J_{1}=2, h=1$ and $K=1000$. We note that for all values of $N$ the concurrence reaches an approximately constant value but then starts oscillating after some critical time $t_c$, that increases as $N$ increases, which means that the oscillation will disappear as we approach an infinite one-dimensional system. Such oscillations are caused by the spin-wave packet propagation \cite{HuangPhysRev}.
We next study the dynamics of the nearest neighbor concurrence when the coupling parameter is $J_{cos}$ with different values of $K$, i.e. different frequencies, which is shown in Fig.~\ref{fig:Jcos}. We first note that $C(i,i+1)$ shows a periodic behavior with the same period of $J(t)$. It has been shown in a previous work \cite{Sadiek_2010} that for the considered system at zero temperature the concurrence depends only on the ratio $J/h$. When $J \approx h$ the concurrence has a maximum value. While when $J>>h$ or $J<<h$ the concurrence vanishes. In Fig.~\ref{fig:Jcos}, one can see that when $J=J_{max}$, $C(i,i+1)$ decreases because large values of $J$ destroy the entanglement, while $C(i,i+1)$ reaches a maximum value when $J=J_{0}=0.5$. As $J(t)$ vanishes, $C(i,i+1)$ decreases because of the magnetic field domination.
In Fig.~\ref{fig:Jsin} we study the dynamics of nearest neighbor concurrence when $J=J_{sin}$. As can be seen, $C(i,i+1)$ shows a periodic behavior with the same period as $J(t)$. We note that we get larger values of $C(i,i+1)$ compared to the previous case $J=J_{cos}$. This indicates the importance of an initial concurrence to maintain and yield high concurrence as time evolves.
Comparing our results with the previous results of time dependent magnetic field \cite{HuangPhysRev}, we note that the behavior of $C(i,i+1)$ when $J=J_{cos}$ is similar to its behavior when $h=h_{sin}$, where $h_{sin}=h_{0}\left(1- \sin\left(K t\right)\right)$, and vice versa.
\clearpage
\subsection{A Time-Dependent Magnetic Field}
\begin{figure}[htbp]
\begin{minipage}[c]{\textwidth}
 \centering 
  	\subfigure[]{\label{fig:bothcos}\includegraphics[width=7cm]{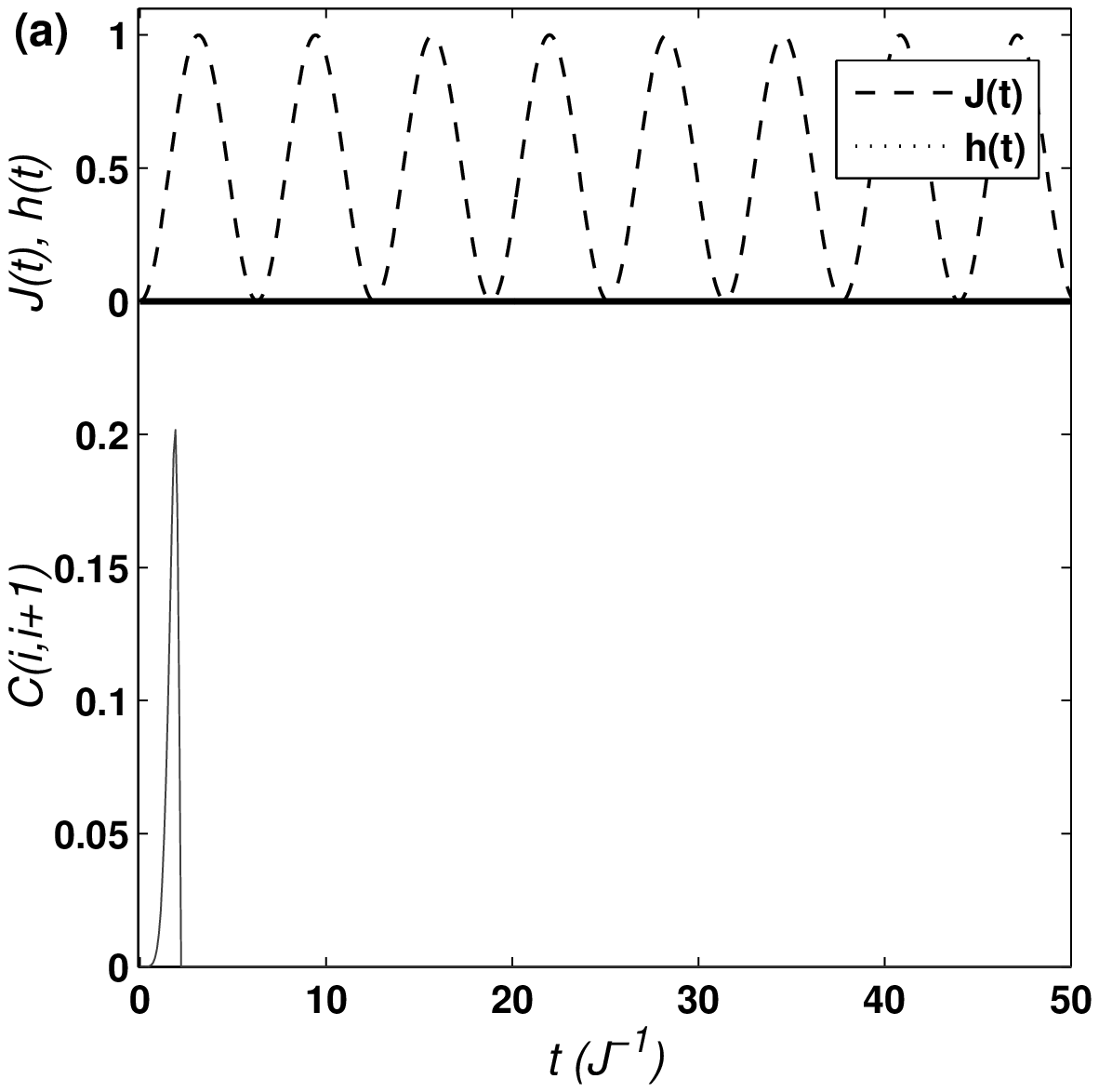}}
  	\subfigure[]{\label{fig:bothsin}\includegraphics[width=7cm]{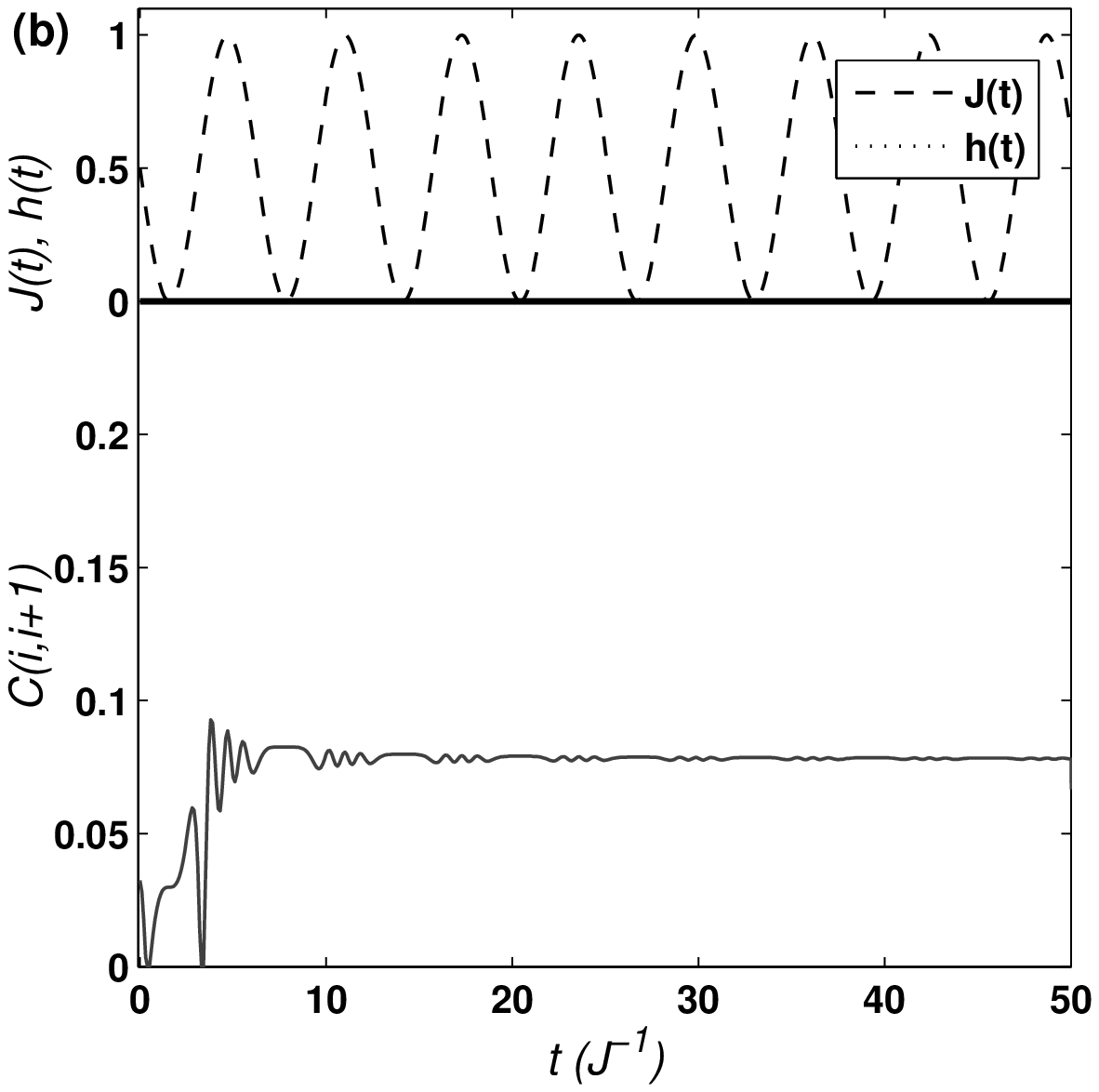}}
		\caption{{\protect\footnotesize Dynamics of nearest neighbor concurrence with $\gamma=1$ at $kT=0$ with $J_{0}=h_{0}=0.5,K=1$ for (a) $J_{cos}$ and $h_{cos}$ ; (b) $J_{sin}$ and $h_{sin}$.}}
  \label{fig:both2}
\end{minipage}
\end{figure}

In this section we use the exact solution to study the concurrence for four forms of coupling parameter $J_{exp}, J_{tanh}, J_{cos}$ and $J_{sin}$ when $J(t)=\lambda h(t)$ where $\lambda$ is a constant. We have compared the exact solution results with the numerical ones and they have shown coincidence.
\begin{figure}[htbp]
\begin{minipage}[c]{\textwidth}
 \centering 
  	\subfigure[]{\label{fig:Jexp}\includegraphics[width=6.5cm]{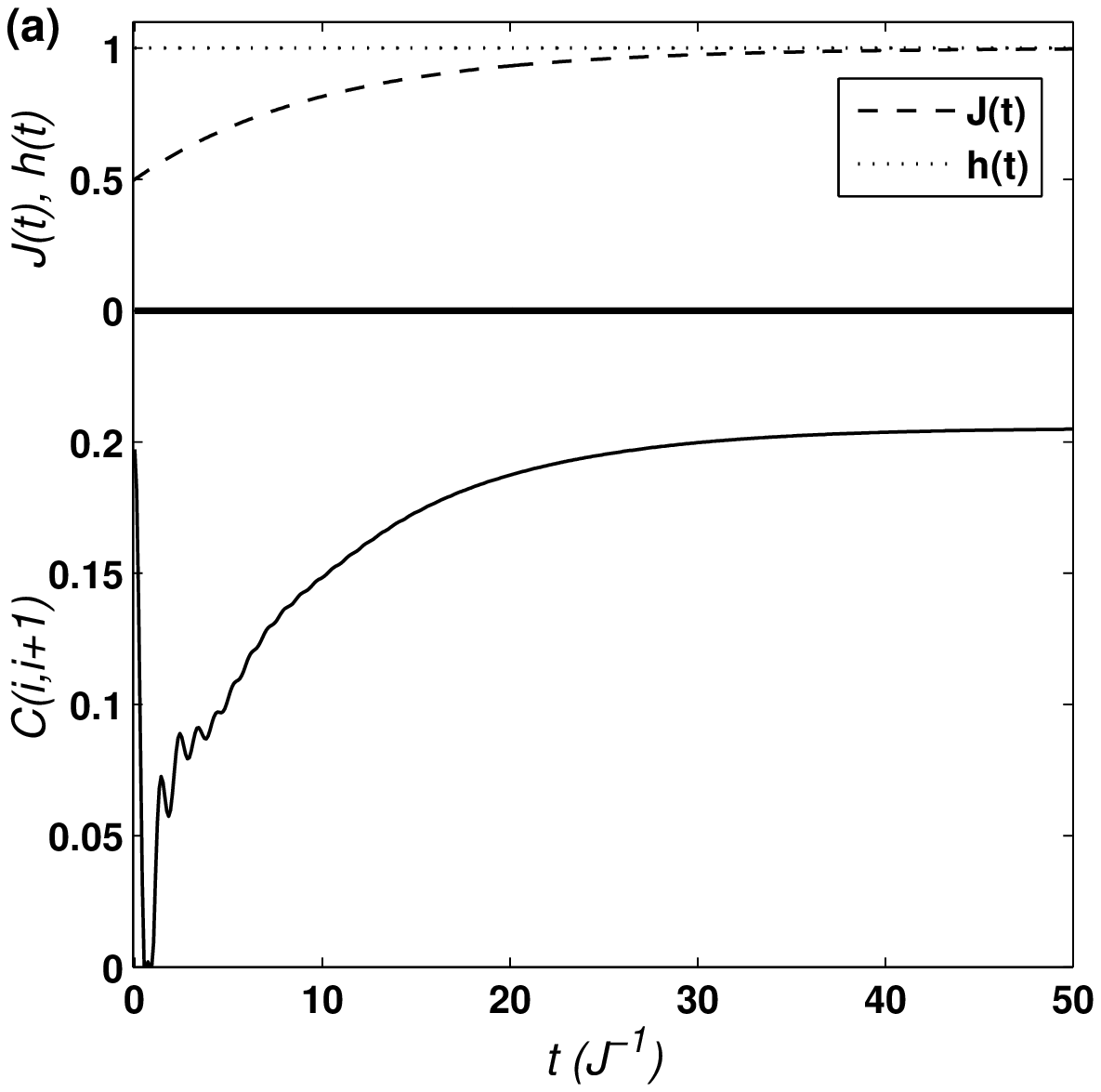}}
  	\subfigure[]{\label{fig:bothexp}\includegraphics[width=6.5cm]{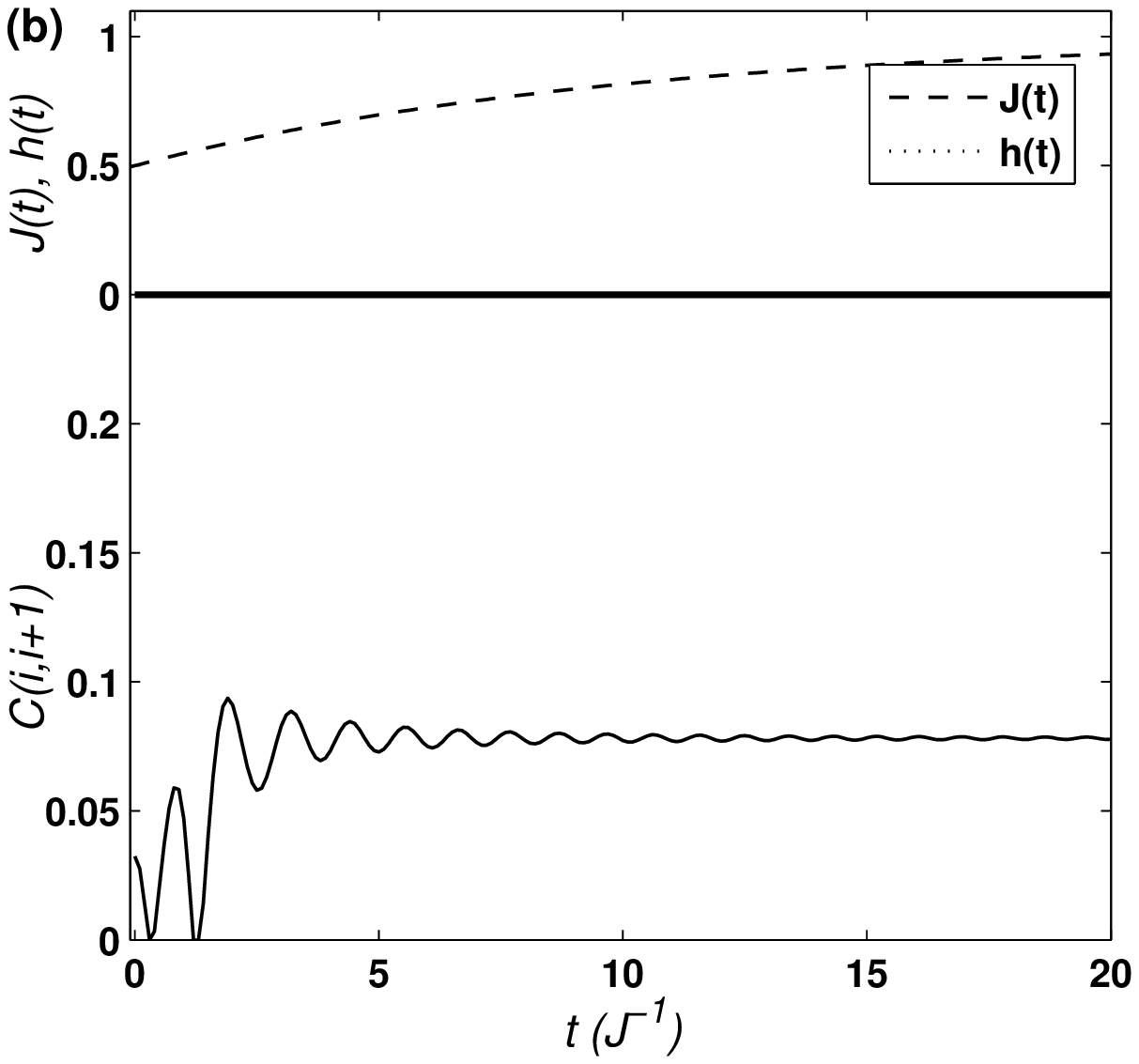}}
  	\subfigure[]{\label{fig:bothtanh}\includegraphics[width=6.5cm]{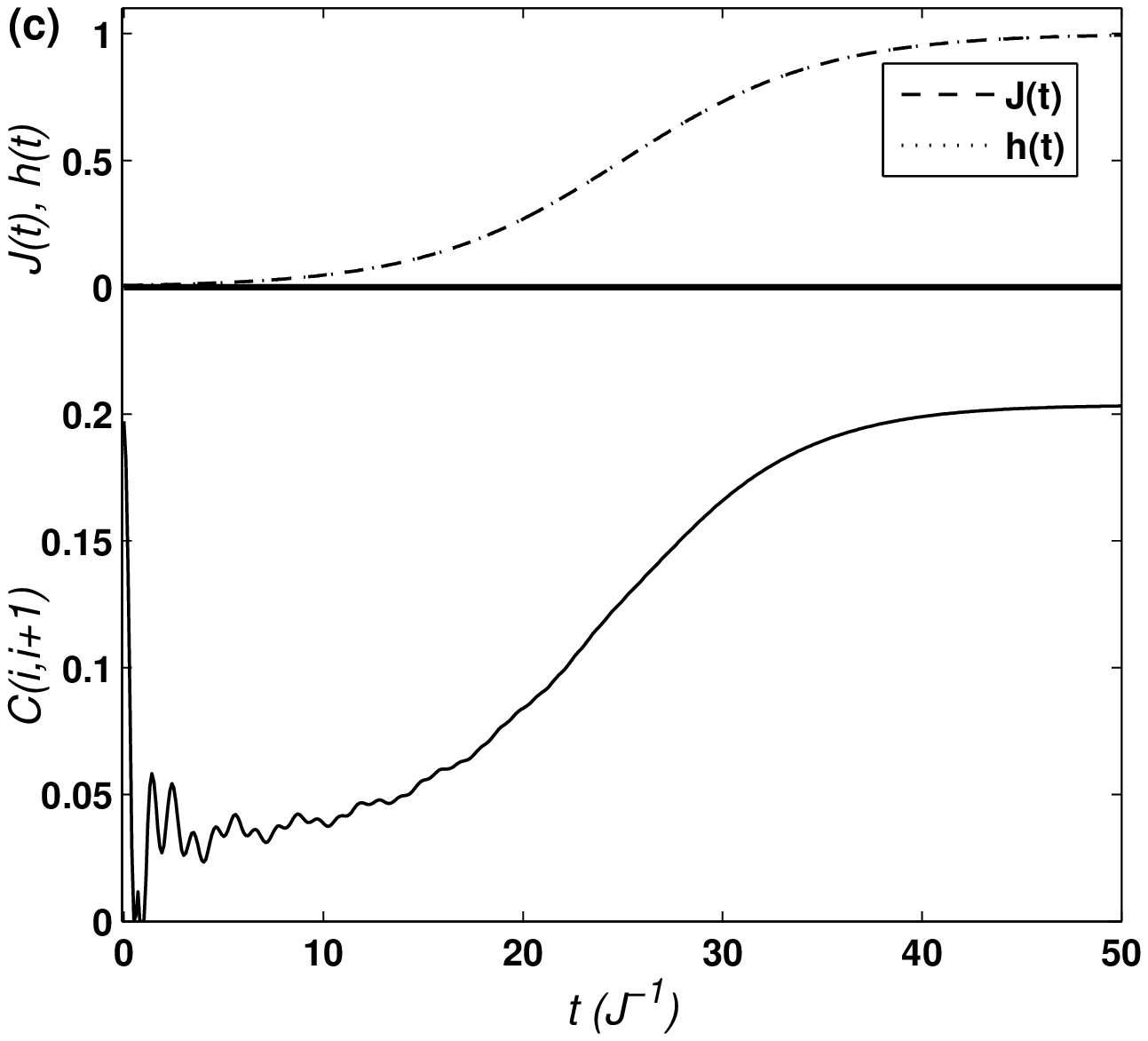}}
  	\caption{{\protect\footnotesize Dynamics of nearest neighbor concurrence with $\gamma=1$ at $kT=0$, $J_{0}=0.5, J_{1}=1, K=0.1$ and (a) $h(t)=1$ ; (b) $h(t)=J(t)=J_{exp}$; (c) $h(t)=J(t)=J_{tanh}$.}}
  \label{fig:both1}
\end{minipage}
\end{figure}
The dynamics of $C(i,i+1)$ for $h(t)=1$ and $J=J_{exp}, J_{0}=0.5, J_{1}=1$ with $K=0.1$ is explored in Fig.~\ref{fig:Jexp}. Comparing with Fig.~\ref{fig:bothexp}, which shows the dynamics of $C(i,i+1)$ for $h(t)=J(t)=J_{exp}, J_{0}=0.5, J_{1}=1$ and $K=0.1$, as one can see the time-dependent magnetic field caused the asymptotic value of $C(i,i+1)$ to decrease. A similar behavior occurs when $h(t)=J(t)=J_{tanh}, J_{0}=0.5, J_{1}=1$ with $K=0.1$ as exploited in Fig.~\ref{fig:bothtanh}. Figures~\ref{fig:both2}(a) and (b) show the dynamics of $C(i,i+1)$ when $h(t)=J(t)=J_{cos}$ and $h(t)=J(t)=J_{sin}$ respectively, where $J_{0}=0.5$ and $K=1$. As can be noticed the concurrence in this case does not show a periodic behavior as it did when $h(t)=1$ in Figs.~\ref{fig:Jcos} and \ref{fig:Jsin}.

\begin{figure}[htbp]
\begin{minipage}[c]{\textwidth}
 \centering 
  	\label{fig:JexpLvar}\includegraphics[width=7cm]{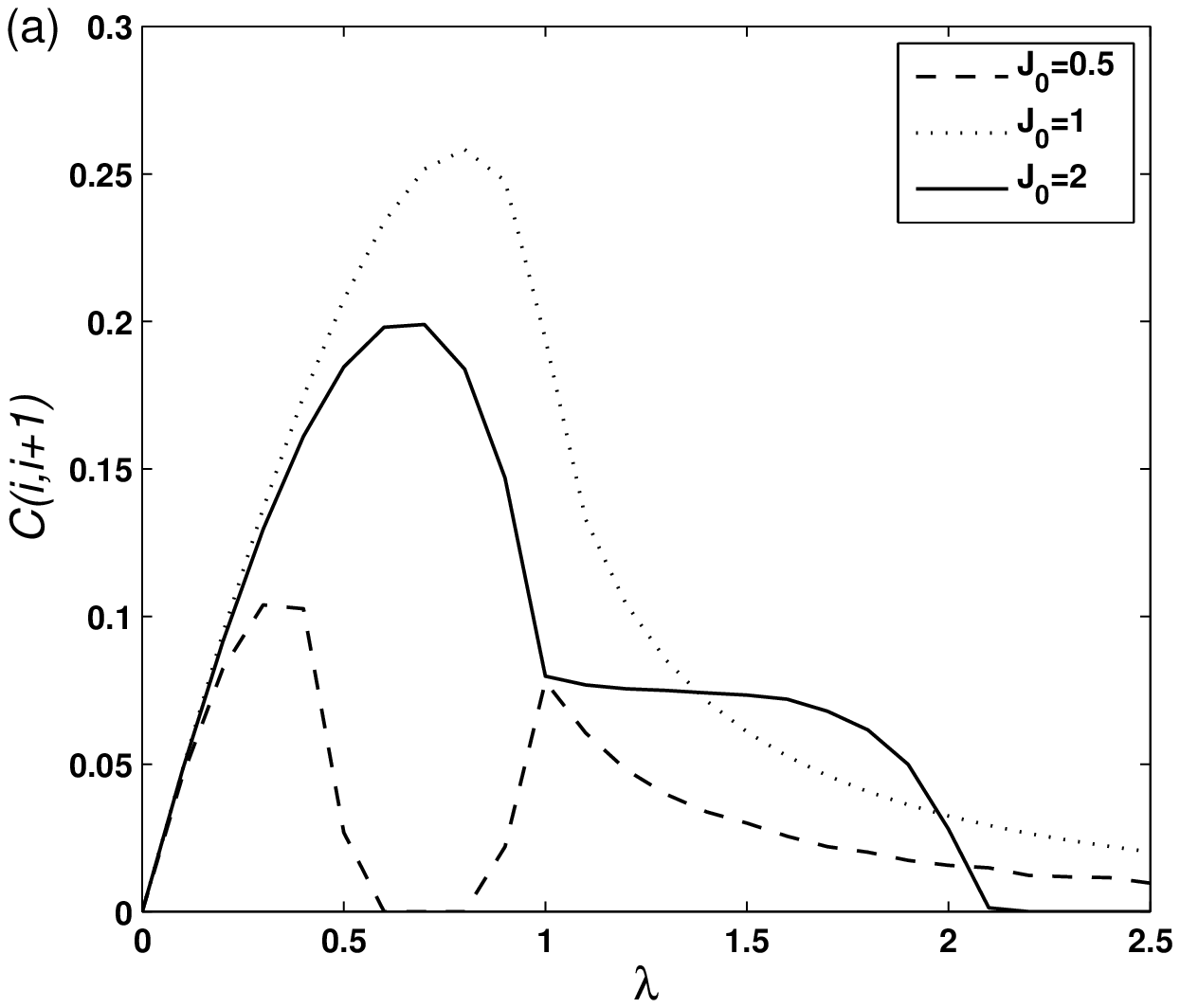}
  	\label{fig:JLvarkTnon0}\includegraphics[width=7cm]{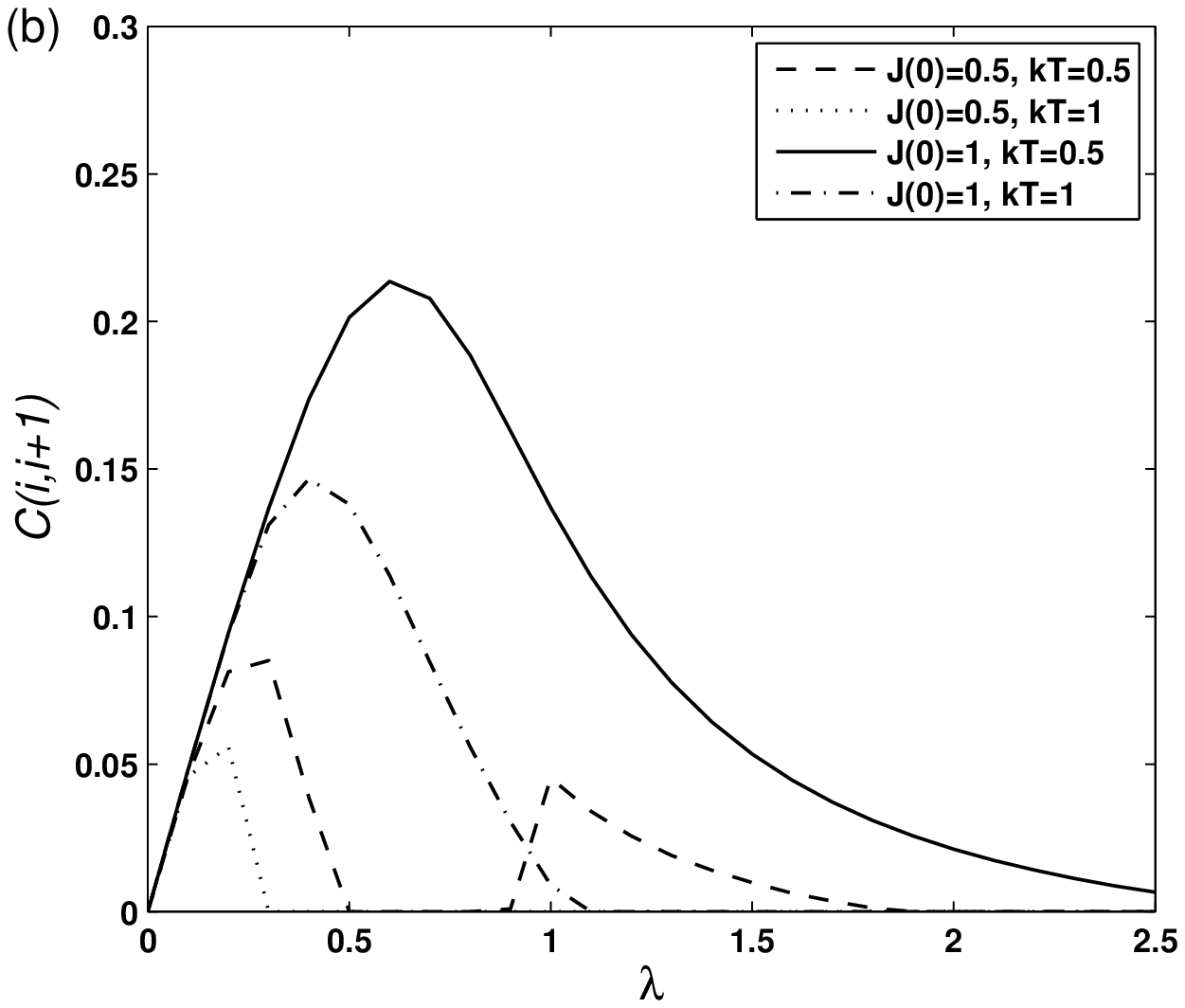}
		\caption{{\protect\footnotesize The behavior asymptotic value of $C(i,i+1)$ as a function of $\lambda$ with $\gamma=1$ at (a) $kT=0$ ; (b) $kT=0.5, 1$.}}
  \label{fig:Lvar}
\end{minipage}
\end{figure}

In Fig.~\ref{fig:Lvar}(a) we study the behavior of the asymptotic value of $C(i,i+1)$ as a function of $\lambda$ at different values of the parameters
$J_0, J_1$ and $K$ where $J(t)=\lambda h(t)$. Interestingly, the asymptotic value of $C(i,i+1)$ depends only on the initial conditions not on the form or behavior of $J(t)$ at $t > 0$. This result demonstrates the sensitivity of the concurrence evolution to its initial value. Testing the concurrence at non-zero temperatures demonstrates that it maintains the same profile but with reduced value with increasing temperature as can be concluded from Fig.~\ref{fig:Lvar}(b). Also the critical value of $\lambda$ at which the concurrence vanishes decreases with increasing temperature as can be observed, which is expected as thermal fluctuations destroy the entanglement.

\begin{figure}[htbp]
\begin{minipage}[c]{\textwidth}
 \centering 
  	\label{fig:Gamma05}\includegraphics[width=7cm]{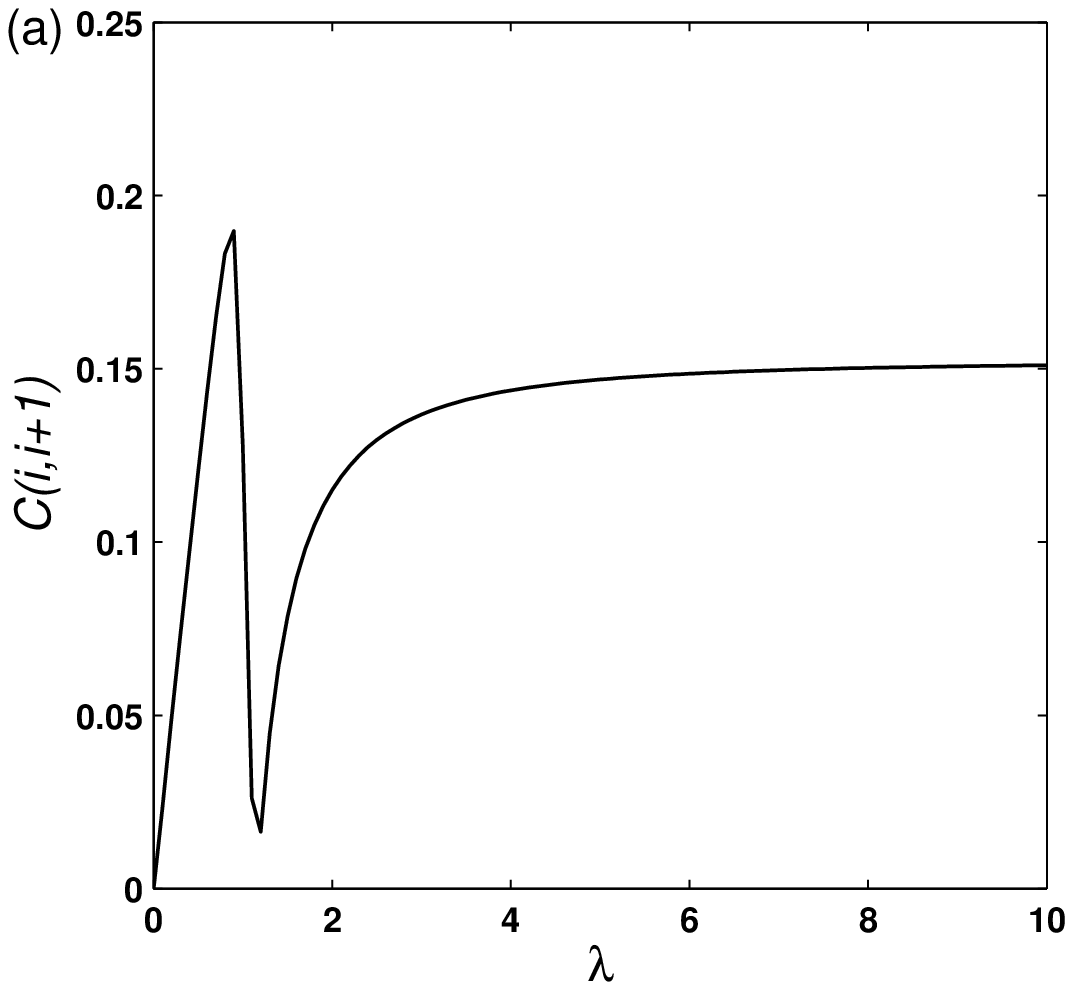}
  	\label{fig:Gamma0}\includegraphics[width=7cm]{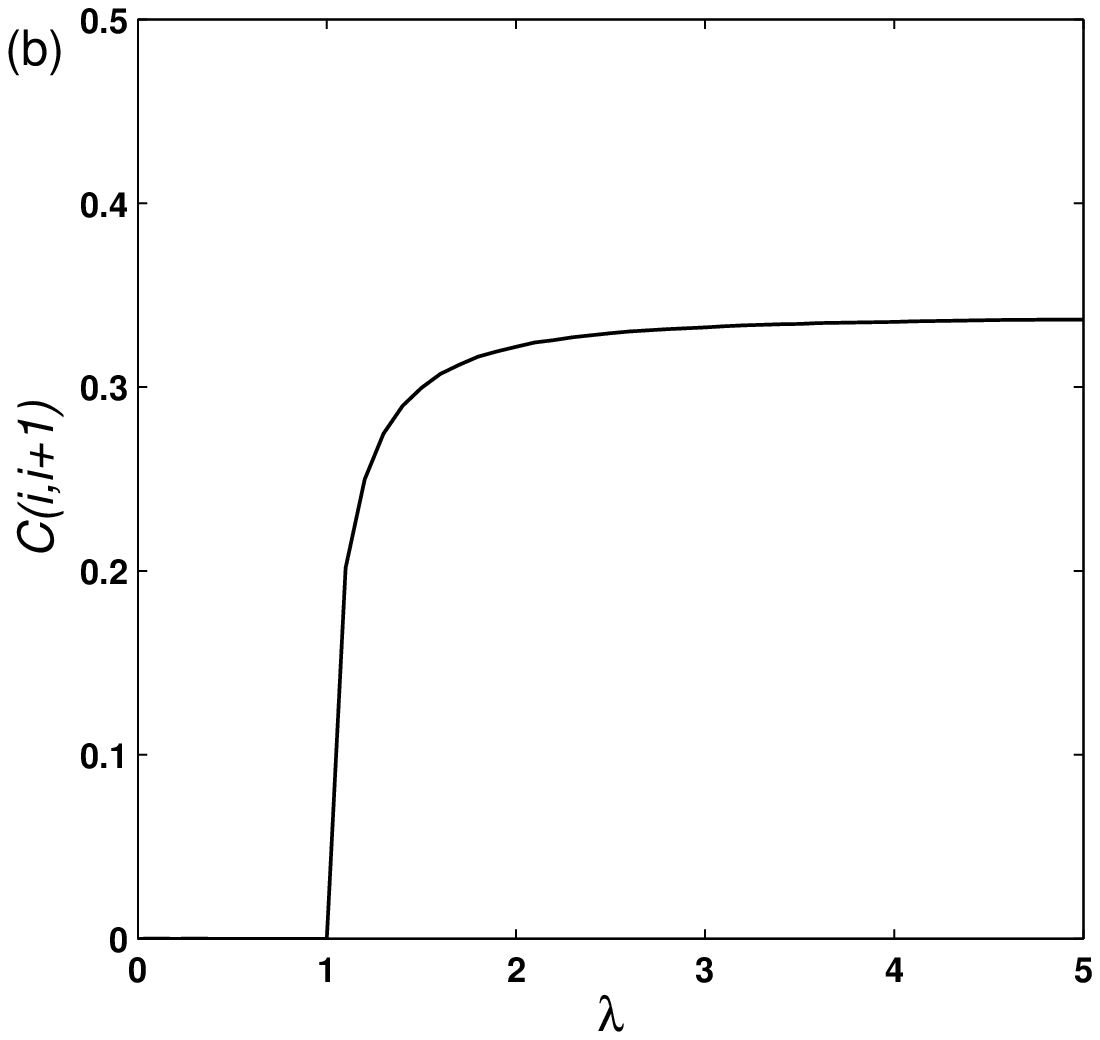}
		\caption{{\protect\footnotesize The behavior asymptotic value of $C(i,i+1)$ as a function of $\lambda$ at $kT=0$ with (a) $\gamma=0.5$ ; (b) $\gamma=0$.}}
  \label{fig:Gamma050}
\end{minipage}
\end{figure}

Finally, in Fig.~\ref{fig:Gamma050} we study the partially anisotropic system, $\gamma=0.5$, and the isotropic system $\gamma=0$ with $J(0)=1$. We note that the behavior of $C(i,i+1)$ in this case is similar to the case of constant coupling parameter studied previously \cite{Sadiek_2010}. We also note that the behavior depends only on the initial coupling $J(0)$ and not on the form of $J(t)$ where different forms have been tested.
\section{Conclusions and Future Directions}
We have studied the dynamics of entanglement in a one-dimensional $XY$ spin chain coupled through a time-dependent nearest neighbor coupling and in the presence of a time-dependent magnetic field at zero and finite temperatures. We presented a numerical solution for the system for general $J(t)$ and $h(t)$ and an exact solution for proportional $J(t)$ and $h(t)$. For an exponentially increasing $J(t)$ we found that the asymptotic value of the concurrence depends on the exponent transition constant value, which confirms the non-ergodic behavior of the system. For a periodic $J(t)$ we found that the concurrence shows a periodic behavior with the same period as $J(t)$. On the other hand for both periodic coupling and magnetic field with same period, the concurrence loses its periodic behavior. When $J(t)=\lambda h(t)$ where $\lambda$ is a constant we found that the asymptotic value of the concurrence depends only on the initial conditions regardless of the form of the coupling parameter or the magnetic field.
In future, we would like to study the effect of an impurity spin on the entanglement along the driven one-dimensional spin chain. It will be also interesting to study the decoherence of a spin pair (quantum gate) as a result of coupling to a driven one-dimensional spin chain acting as its environment.
\section*{Acknowledgments}

This work was supported in part by the deanship of scientific research, King Saud University.

\end{document}